# The science of fake news

David Lazer[1,2*], Matthew Baum[2*], Yochai Benkler[2], Adam Berinsky[3], Kelly Greenhill[4,2], Filippo Menczer[5], Miriam Metzger[6], Brendan Nyhan[7], Gordon Pennycook[8], David Rothschild[9], Michael Schudson[10], Steven Sloman[11], Cass Sunstein[2], Emily Thorson[12], Duncan Watts[9], Jonathan Zittrain[2]

[1]Northeastern University, [2]Harvard University, [3]MIT, [4]Tufts University, [5]Indiana University--Bloomington, [6]University of California--Santa Barbara, [7]Dartmouth University, [8]Yale University, [9]Microsoft Research, [10]Columbia University, [11]Brown University, [12]Boston College

*contributed equally to this article.

**One Sentence Summary**: Addressing fake news requires a multidisciplinary effort to understand how the Internet spreads content and how people process news.

**Abstract:** Fake news emerged as an apparent global problem during the 2016 U.S. Presidential election. Addressing it requires a multidisciplinary effort to define the nature and extent of the problem, detect fake news in real time, and mitigate its potentially harmful effects. This will require a better understanding of how the Internet spreads content, how people process news, and how the two interact.  We review the state of knowledge in these areas and discuss two broad potential mitigation strategies: better enabling individuals to identify fake news, and intervention within the platforms to reduce the attention given to fake news. The cooperation of Internet platforms (especially Facebook, Google, and Twitter) with researchers will be critical to understanding the scale of the issue and the effectiveness of possible interventions.

**Main Text:**

The rise of fake news highlights an urgent need for a multidisciplinary scientific effort to study misinformation in the emergent news ecosystem of the 21st century. Longstanding institutional bulwarks against misinformation have proven inadequate in the Internet age. Scientific research should inform the societal challenge of building a new system of safeguards. First, we need to define the nature and extent of the problem. How prevalent is fake news, and how much influence does it have on people? What form does that influence take? Second, we need to determine whether we can intervene to reduce the scale and impact of fake news. Can we empower individuals to detect and ignore it, through fact checking and education? What actions can platforms take to reduce its flow? Can fake news and social manipulations be algorithmically detected? How might industry collaborate with academics to identify answers to these questions? This essay offers preliminary assessments of these foundational questions.

DEFINITION

We define fake news to be fabricated information that mimics the output of the news media in form, but not in organizational process or intent. Most notably, fake news outlets lack the news media's editorial norms and processes for ensuring the accuracy and credibility of information. Fake news is a subcategory of misinformation—incorrect or misleading information about the state of the world. It is particularly pernicious in that it undermines the credibility of standard news outlets. We recognize that some have advocated eschewing the phrase because of its use as a political weapon. We have retained it both because of its value as a distinct scientific construct, and because its political salience helps draw attention to an important subject. Below, we offer some preliminary guideposts for converting this conceptual definition into an operational research program.

BACKGROUND

The ancien régime protecting against misinformation resulted from the emergence of journalistic norms of objectivity and balance that arose as a backlash among journalists against the widespread use of propaganda in World War I—particularly the role journalists had played in propagating it—as well as against the rise of corporate public relations in the 1920s (for detailed supporting literature on this and other points, see SM). Local and national oligopolies created by the dominant technologies of information distribution in the 20th century (publishing and broadcasting) sustained these norms. The Internet has removed many of those constraints on dissemination, contributing to the abandonment of traditional news sources that had long enjoyed high levels of public trust and credibility. For instance, U.S. newspaper circulation fell 30% from 1990 to 2012 *(1)*, while audience ratings for network television evening news programs plummeted by 56% from 1980 to 2010 *(2)*. In parallel, general trust in the mass media collapsed, especially on the political right. In 1997, 64% of Democrats and 41% of Republicans reported a great deal or fair amount of trust in the media. Twenty years later, in 2016, these percentages had fallen by 13 percentage points (to 51%) among Democrats, and by 27 points (to 14%) among Republicans *(3)*.

The United States has experienced an important evolution in its geo/socio/political environment alongside these changes in the news ecosystem. Geographic polarization of partisan preferences has dramatically increased over the last 40 years *(4-5)*, reducing opportunities for cross-cutting political interaction. Homogenous social networks, in turn, reduce our tolerance for alternative points of view, amplify attitudinal polarization, boost our likelihood of accepting ideologically compatible news, and increase closure to new information.

We believe these trends have created a context in which fake news can attract a mass audience. Below we discuss the social science and computer science research regarding belief in fake news and the mechanisms by which it spreads. We focus particular attention on the many unanswered scientific questions raised by the recent proliferation of politically-oriented fake news.

PREVALENCE AND IMPACT

How large a problem is fake news? How common is it, and what impact does it have on individuals? There are surprisingly few scientific answers to these basic questions; and any efforts devoted to possible solutions should be commensurate to the magnitude of the societal problem. In evaluating prevalence of fake news, we advocate focusing on publishers rather than individual stories, because we view the defining element of fake news to be the intent and processes of the publisher. This has the additional advantage of avoiding the morass of evaluating the "fakeness" of every single news story.

One study evaluating the dissemination of a set of prominent fake news stories estimated that the average American encountered between one and three fake news stories during the month before the election *(6)*. This likely is a conservative estimate, since the study tracked only 156 fake news stories. We do know that, as with legitimate news, many fake news stories have gone viral on social media, suggesting that many people are exposed. However, knowing how many social media accounts encountered or even shared a piece of fake news is not the same as knowing how many people read or were affected by it. Metrics such as sharing and liking are subject to the same manipulations that enable fake news. For instance, the impact of a fake news story that has been shared by millions of automated accounts but only a few humans is quite different from the opposite. Despite recent tools to track the spread of fake news, the scientific capacity to measure human attention to identified fake news content on the Internet is still limited, except in a lab setting (discussed below).

Importantly, exposure does not equal impact. Evaluation of the medium-to-long-run impact of exposure to fake news on political behavior (e.g., whether and how to vote) is essentially nonexistent in the literature. Beyond political impacts, what we know about media effects more generally suggests many potential pathways of influence, from increasing cynicism and apathy to encouraging extremism. There is, however, little evaluation of the impact of fake news in these regards. The scientific community thus needs to develop collective resources for evaluating how fake (and real) news exposure affects actual people, with known attributes and opinions. There are some proprietary panels aimed at commercial markets devoted to studying people's attention on the web; there need to be equivalent resources that are widely available for scientific research.

INTERVENTIONS: EMPOWERING THE INDIVIDUAL, CHANGING THE STRUCTURE

What interventions might be effective at stemming the flow of fake news? We identify two categories of interventions: (1) those aimed at empowering individuals to evaluate the fake news they encounter (in particular, fact checking and education); and (2) structural changes aimed at preventing exposure of individuals to fake news in the first instance (i.e., changes in the policies of the platforms). For present purposes, we focus on what we know and need to know regarding the effectiveness of potential interventions.

EMPOWERING INDIVIDUALS

Currently, one of the primary interventions focused on stemming the flow of fake news is fact checking: the direct identification of content that is factually incorrect. There are many forms of fact checking, from stand-alone websites that evaluate the factual claims of news reports, such as Politifact and Snopes, to evaluations of news reports by credible news media, such as the Washington Post, to warnings on content placed by informational intermediaries, such as Facebook.

Despite the apparent elegance of fact checking, the science supporting its efficacy as a solution to fake news is at best mixed. Research shows that people more often use the media for personal gratification than for truth seeking. Additional research demonstrates that people prefer information that confirms their pre-existing attitudes (selective exposure) and view information consistent with their pre-existing beliefs as more persuasive than dissonant information (confirmation bias). That is, prior partisan and ideological beliefs might prevent acceptance of fact checking of a given fake news story. Ironically, this means that those most likely to be deceived by fake news are least likely to believe any attempt to prevent their own deception.

To make matters worse, the way that human memory operates means that unless conducted carefully, fact checking might even be counterproductive under certain circumstances. Research on fluency—the ease of information recall—and familiarity bias in politics shows that people tend to remember information, or how they feel about it, while forgetting the context within which they encountered it. They are also more likely to accept familiar information as true. Thus, providing any information may increase an individual's likelihood of accepting it as true when encountered again. The literature presents contradictory evidence about the effectiveness of claim repetition in fact checking. Experimental and survey research has confirmed the basic familiarity effect, though recent experiments suggest that retractions are more effective in reducing misinformation effects when they explicitly repeat the misinformation. Further research is needed to reconcile these contradictions and determine the conditions under which fact checking interventions are most effective.

A second, longer-run, approach seeks to improve individual evaluation of the quality of information sources through education. There has been a recent proliferation of efforts to inject training of critical information skills in primary and secondary schools. However, we have found no research on evaluating the causal impact of general training in critical information consumption skills on accurate assessments of news source credibility. Moreover, there is an additional risk that an emphasis on fake news might have the unintended consequence of reducing the perceived credibility of real news outlets. Given the array of potential approaches, and the possibility of undermining trust in news generally, there is a great need for a rigorous program evaluation of different educational interventions.

In summary, the evidence that intervention at the individual level—through either fact checking or teaching critical skills—will solve the problem of fake news is at best quite limited. This may reflect broader tendencies in collective cognition, as well as structural changes in our society. Typically, individuals evaluate the credibility of information they encounter only if it violates their preconceptions or they are incentivized to do so. Otherwise, they tend to accept information uncritically. People also tend to align their beliefs with the values of their community. Geographic polarization along political lines can thus facilitate the emergence of self-reinforcing alternative realities within a group. Policies of Internet platforms that rank order the delivery of information according to people's preferences may amplify the perception of political homogeneity among one's peers. This set of issues points to the possibility of Internet platform-based interventions aimed at increasing the quality and diversity of the information that flows into a community.

PLATFORM-BASED DETECTION AND INTERVENTION: ALGORITHMS AND BOTS

Internet platforms have become the most important enablers and primary conduits of fake news. It is, for instance, inexpensive to create a website that has the trappings of a professional news organization. It has also been easy to monetize its content through online ads and social media dissemination. The Internet not only provides a platform for publishing (mis)information, but is a networked medium that actively promotes its dissemination.

About 44% of Americans overall report getting news from social media often or sometimes, with Facebook as by far the dominant source *(7)*. Social media are far more important than mainstream media as conduits for fake news sites *(6)*. Russia successfully manipulated all of the major platforms during the 2016 election, as their recent testimony before the Senate Judiciary Committee made clear *(8)*. Is it possible to rewire the Internet to reduce the spread and impact of fake news? The point of intervention would be the big platform companies, most notably

Google, Facebook, and Twitter. They are often the mediators of not only our relationship with the news media, but also with our friends and relatives. Generally, their business model relies on monetizing attention through advertising. They use complex statistical models to predict and maximize engagement with content. For example, most Facebook users do not see all of the content produced by their Facebook friends, but just the content with which Facebook predicts they are likely to engage.

It is quite plausible that the emphasis by search and social media platforms on optimizing attention from consumers increases selective exposure. Google or Facebook may provide users with slanted content because they predict that someone like them is more likely to engage with it. There is an emerging literature on the algorithmic underpinnings of the 21$^{st}$ century world, including some evidence of this amplification of selective exposure of political content for Facebook and Google. However, even though this research is fairly recent, it represents rapidly fading snapshots of a moment in history. There exists little research focused specifically on fake news and no comprehensive data collection system to provide a dynamic understanding of how these pervasive systems are evolving. It is impossible to recreate the Google of 2010; Google itself would be unable to do this even if they had a record of the underlying code, because the patterns emerge from a complex interaction amongst code, content, and users. However, it is possible to thoroughly capture for posterity what the Google of 2017 is doing. More generally, researchers need to conduct a rigorous ongoing audit of how the major platforms filter information.

There is ample evidence that platforms are highly vulnerable to manipulation. By liking, sharing, and searching for information, social bots (automated accounts impersonating humans) and extreme partisans can amplify the reach of fake news. Bots are numerous and commoditized. They were responsible for a significant portion of political content posted during the 2016 campaign, and some of the same bots later attempted to influence the 2017 French election *(9)*. These tactics also aim to manipulate the algorithms that platforms utilize to predict potential engagement with content by a wider population. Indeed, a recent Facebook white paper reports widespread efforts to carry out this sort of manipulation during the 2016 election *(10)*.

The discussion above suggests various possible interventions by platforms. Consumers could be provided signals of source quality. Source quality could be incorporated into the algorithmic rankings of content, and personalization of political information could be minimized relative to other types of content. Functions that emphasize currently trending content could seek to exclude bot activity from measures of what is trending. More generally, the automated spread of news content by bots and cyborgs (humans aided by automated posting to their accounts) could be curbed.

The platforms have surely attempted each of these steps, in some form, and many others. They have posted statements about their efforts, as well as reports, such as Facebook's aforementioned white paper on information operations that claimed that manipulations by malicious actors accounted for less than one tenth of 1% of civic content shared on the platform. However, these and other such claims are unverifiable. Indeed, researchers, politicians, and the public should dismiss as self-serving any claims that the platforms make regarding addressing the problem of misinformation that cannot be verified by third parties.

This points to the value for platforms of collaborating with academics on evaluations of the scope of the issue and the effectiveness of their interventions There are multiple challenges to scientific collaboration, both from the perspective of industry and academics; however, these barriers have been tackled in other contexts. We thus believe these issues are surmountable, and that there is an ethical responsibility for the platforms to collaborate on the science of fake news.

The possible effectiveness of platform-based policies suggests either self-regulation by the platforms or government intervention. Direct government regulation of an area as sensitive as news carries its own risks, such as, for instance, whether or not government regulators could maintain (and, as important, be *seen* as maintaining) impartiality in enforcing any content-based distinctions. An alternative to direct government regulation would be to enable tort lawsuits alleging, for example, defamation.  To the extent an online platform assisted in the spreading of a manifestly false (but still persuasive) story, there might be avenues for liability, which in turn would pressure platforms to intervene more regularly. In the U.S. context, however, a provision of the 1996 Communications Decency Act offers near-comprehensive immunity on this front to platforms for false or otherwise-actionable statements penned by others. Any change to this legal regime, however, itself raises thorny issues about the extent to which platform content (and content curation) should be subject to second-guessing by people alleging injury.

Structural interventions generally raise legitimate concerns about respecting human agency. But just as the media oligopolies of the 20$^{th}$ century shaped the information to which Americans were exposed, the far-more-vast Internet oligopolies are already shaping human experience on a global scale. The question before us is how those immense powers are being—and should be—exercised.

CONCLUSION

Current hand-wringing over fake news in many ways mirrors widespread concern in the mid-20th century that propaganda—by Nazis, and later by Communists—posed a fundamental threat to democracy. These concerns prompted the post-World War II generation of sociologists, psychologists, and political scientists to pioneer the social-psychological school of political behavior research. Ironically, these scholars found little evidence of such a threat. Yet subsequent generations of scholars, building on these foundational works, have found that even if the media (the presumed purveyors of propaganda) typically cannot change what we think (that is, our attitudes), they are far more successful at changing what we think *about* (via priming, framing, and agenda setting). The analogy to fake news is clear: even if fake news does not alter most people's beliefs, it may nonetheless reinforce existing false beliefs, increase their salience, or shape the news agenda, with potentially harmful effects for society. Explicating and countering such effects requires a multidisciplinary research program similar to the post-World War II scholarly effort aimed at countering propaganda effects.

Our call here is to promote interdisciplinary research spanning psychology, computer science, political science, economics, law, and communication with the normative objective of reducing the spread of fake news and of addressing the underlying pathologies it has revealed. The failures of the news in the early 20th century led to the rise of a set of journalistic norms and practices that, while imperfect, generally served us well by striving to provide objective, credible information. We must again redesign our news ecosystem in the 21st century. Doing so will require new, multidisciplinary science on fake news and misinformation.